\documentstyle{article}

\font\black=msbm10

\def\sp #1{{{\cal #1}}}

\def\field #1{\hbox{{\black #1}}}

\def\lie #1{{\sp L_{\!#1}}}    					
\def\pd#1#2{\frac{\partial#1}{\partial#2}}
           			
\def\set#1{\{\,#1\,\}}            
\def\>#1{{\bf #1}}               
\def\dd#1{\frac{\partial}{\partial#1}}

\def\Eq#1{{\begin{equation} #1 \end{equation}}}
           				
\def\mitad{\frac{1}{2}}

\def\R{{\hbox{{\field R}}}}

\def\T{{\hbox{{\field T}}}}
						
\def\dim{\hbox{{\rm dim}}} 					
   
\def\rank{{\hbox{\rm rank}}}     		
\def\corank{{\hbox{\rm corank}}}   
\def\span{{\hbox{\rm span\,}}}  		

\newtheorem{theorem}{{\sc Theorem}}

\begin{document}

{\mbox{   }}

\vskip 2cm

\centerline{\bf A. Ibort, M. de Le\'on, G. Marmo, D. Mart{\'\i}n de Diego}  

\vskip 1cm

\centerline{\large\bf NON--HOLONOMIC CONSTRAINED SYSTEMS} 

\vskip 0.25cm

\centerline{\large\bf AS IMPLICIT DIFFERENTIAL EQUATIONS}

\vskip 2.5cm

{\parindent 0cm{\small {\bf Abstract.} Non-holonomic constraints, both in the
Lagragian and Hamiltonian formalism, are discussed from the geometrical 
viewpoint 
of implicit differential equations.   A precise statement of both problems is
presented remarking the similarities and differences with other classical
problems with constraints.   In our discussion, apart from a constraint
submanifold, a field of permitted directions and a system of reaction forces are
given, the later being in principle unrelated to the constraint submanifold.   
An implicit differential equation is associated to a
non-holonomic problem using the Tulczyjew's geometrical description of
the Legendre transformation.  The integrable part of this implicit differential
equation is extracted using an adapted version of the integrability algorithm. 
Moreover, sufficient conditions are found that guarantees the compatibility of 
the
non-holonomic problem, i.e., that assures that the integrability algorithm stops
at first step, and moreover it implies the existence of a vector field whose
integral curves are the solutions to the problem.  In addition this vector field
turns out to be a second order differential equation.  These compatibility
conditions are shown to include as particular cases many others obtained
previously by other authors.  Several examples and further lines of development
of the subject are also discussed. }}

\bigskip
\bigskip

\section{Introduction}

Non-holonomic constraints have been the subject of deep 
analysis (not exent from some controversy) since the dawn of Analitical 
Mechanics. 
In fact D'Alembert's principle of virtual work \cite{Da43}, \cite{La88}, and 
Gauss
principle of least constraint \cite{Ga29} can be considered to be the first
solutions to the analysis of systems with constraints, holonomic or not.   A
golden age for the subject came with the contributions of O. H\"older 
\cite{Ho96},
G. Hamel \cite{Ha04}, P. Appell
\cite{Ap11}, E. Delassus \cite{De11}, T. Levi--Civita \cite{Le27}, N. G. Chetaev
\cite{Ch32}, etc., when the discussion of (linear) non-holonomic constraints in
Lagrangian mechanics was considered sistematically.  After the quantum 
revolution
this classical problem was kept frozen in the limbo of the ``postponed''
problems.   However, there has been papers that from time to time have addressed
some of the weak points of the discussions as they were left in the thirties, 
for
instance, the existence and uniqueness of solutions, the inadequacy of Chetaev's
conditions, etc.  Some contributions in this transition era can be found in
the papers by R.J. Eden \cite{Ed51}, V. Valcovici \cite{Va58}, J. Neimark and N.
Fufaev \cite{Ne67}, R. van Dooren \cite{Do75}, V. Rumiantsev \cite{Ru78}, Y.
Pironneau \cite{Pi82}, etc.  

Almost simultaneously with this period a quite revolution was taking place
regarding the foundations of the old discipline of Mechanics.   Geometry was 
used
in a sistematic way to set up its foundations and new and old ideas from 
Geometry
and Mechanics were fusioning in an harmonic and pleasant picture
that has been evolving until today and is not yet completely finished.  
W. Tulczyjew \cite{Tu68} was one of the pionneers of this Geometric revolution
and his ideas and insight have inspired many developments in the
field as is reflected for instance by the variety of contributions to
these Proceedings.   
Sooner or later the geometrical shock wave had to reach also the
problem of non-holonomic constraints. 
In fact, the first geometrical description of
non-holonomic constraints took place in relatively early times in a paper by L.
D. Faddeev and A. M. Vershik \cite{Fa72}.  The importance of this paper was
not recognized until quite recently even if it contained the
first general result on the existence of dynamics of Lagrangian systems with
(not necessarily linear) non-holonomic constraints.  Even previous
to this, J. Klein had already addressed the problem of
constraints using the geometry of Lagrangian systems
\cite{Kl62}.  Much recent are the papers by E. Massa {\it et al}
discussing the geometrical meaning of Chetaev's conditions
\cite{Ma91}; G. Giachetta \cite{Gi92} and M. de Le\'on et al \cite{Let} using 
jet
bundle geometry techniques; J. Koiller \cite{Ko92}, L. Bates {\it et al}
\cite{Ba93} and Bloch {\it et al} \cite{Bl96}  discussing non-holonomic
constraints with symmetry.  More contributions can be found in the papers by L.
Cushmann {\it et al} \cite{Cu92} using a symplectic splitting to describe the
dynamics in presence of non-holonomic constraints; J. Cari\~nena \& M. F.
Ra\~nada \cite{Ca93} geometrizing Lagrange's multipliers; R.
Weber \cite{We86}, A. Van der Shaft {\it et al}
\cite{Va94}, P. Dazord \cite{Da94} and C. M.--Marle \cite{Ma90} addressing the
problem from the Hamitonian viewpoint; Sarlet {\it et al} \cite{Sa95} using
connection theory; de Le\'on {\it et al} \cite{Le96} using almost product
structures and projectors, F. Barone {\it et al} \cite{Ba96} using Tulczyjew's
ideas to set the frame for generalized Lagrangian systems, F. Cardin {\it et al}
\cite{Cr96} combining the vakonomic and non-holonomic approach, and finally
C.M.--Marle in these Proceedings revisiting Faddeev-Vershik conditions.

Our modest contribution to this old subject will consist in
using Tulczyjew's idea of modelling mechanical systems as
implicit differential equations \cite{Me78}, \cite{Ma92}, \cite{Me95}, to 
discuss
non-holonomic constraints.   This approach was already taken by S. Benenti
\cite{Be87} who was able to obtain a set of sufficient conditions for the
existence of dynamics for linear non-holonomic constraints and apply it
successfully to solve the problem of constrained geodesics.  In our approach we
will compare first the problem of Lagrangian systems with non-holonomic
constraints with the problem of constrained Hamiltonian systems.   We will
set up the geometry of Lagrangian and Hamiltonian systems with non-holonomic
constraints and after a brief discussion of implicit differential equations and
the integrability algorithm, we will apply it to the non-holonomic problem.   We
will find that a small modification of the integrability algorithm is needed to
encompass the restrictions imposed by non-holonomic constraints to the solutions
of the problem.  From the analysis of the adapted integrability algorithm some
particular cases arise immediately.   The simplest non-trivial case is discussed
thoroughly and a general condition for the existence of a solution, that turns
out to be a vector field, is discussed.  This geometrical condition contains
as particular cases Faddeev-Vershik's condition, Chetaev's conditions, Bates'
regularity condition, Benenti's conditions, Marle's conditions, etc.  

We will apply these ideas to discuss a several examples that are inspired in
old models like Appell's machine, etc.

\section{Non-holonomic constrained systems}

The discussion to follow of non-holonomic constraints will be set in the realm
of Tulczyjew's triple \cite{Tu1},
\Eq{\label{triple}  T^* (TQ) \stackrel{\alpha}{\longleftarrow} T(T^*Q)
\stackrel{\beta}{\longrightarrow} T^* (T^*Q) .  }
We will recall that $\alpha$ is the Tulczyjew's canonical symplectomorphism from
$T (T^*Q)$ (with its canonical symplectic structure $\dot{\omega}_Q$) to 
$T^*(TQ)$
(with its canonical structure $\omega_{TQ}$), and $\beta$ is the canonical
symplectomorphism defined by the former symplectic structure. 

The problem of constrained Hamiltonian sytems consists in determining
the equations of motion for a system specified by a constraint
submanifold $C\subset T^* Q$ and a Hamiltonian function $H\colon C
\to \R$.   The submanifold $C$ is usually, but not always, determined by a
self-consistency condition of the system under study and it is often the result 
of
a Dirac's type constraint algorithm \cite{Di64}, \cite{Gr92}.   The submanifold 
\Eq{\label{dirac}  D = \set{ v \in T_C (T^*Q) \mid \omega_Q (v,u) =
\langle dH , u \rangle , ~~ \forall u \in TC }  ,}
defines an implicit differential equation on $T (T^*Q)$, called a Dirac system,
whose analysis provide the solution of the posed problem.  Such analysis was
succesfully done in \cite{Me78} and a set of necessary and sufficient conditions
for its integrability was given in the following theorem.

\begin{theorem}
The Dirac system $D$ defined by a constrained Hamiltonian system is
integrable iff $C$ is coisotropic and $H$ projects along the
characteristic distribution of $C$.
\end{theorem}

\bigskip

Despite the similarities between the Hamiltonian constrained problem and the
problem of non-holonomic constraints, the latter has a different nature. 
In the non-holonomic Lagrangian problem a
Lagrangian $L$ will be given in all $TQ$ that defines the unconstrained or
``free'' system.  In fact, we can assume in what follows that $L$ is regular.  
This means that on $TQ$ we have a dynamical vector field given by the
Euler--Lagrange vector field
$\Gamma_L$ defined by the equation,
\Eq{\label{euler_lag} \lie{\Gamma_L} \Theta_L = dL ,}
or equivantly ($L$ regular),
$$i_{\Gamma_L} \omega_L = dE_L ,$$
where $\Theta_L = S^* (dL)$ is the Poincar\'e--Cartan 1--form
defined by $L$, $\omega_L = -d \Theta_L$ is the Cartan 2--form and
$E_L = \Delta (L) - L$ is the energy of $L$, with $\Delta$ the
Liouville vector field on $TQ$.

The canonical tensor field $S$ on $TQ$ is defined in local natural coordinates
$(q^i, v^i)$ on $TQ$ by
$$ S = \dd{v^i} \otimes dq^i ,$$ 
and it allows us to
characterize {\sc sode}'s as $S (\Gamma ) = \Delta$.  The kernel and image of
$S$ consists of vertical vector fields.   On the other hand $S$ acts by duality
on forms and the kernel and image of $S^*$ consists on horizontal 1--forms.
A remarkable property of $S$ and $\omega_L$ is given by the formula,
$$ i_S \omega_L = 0 ,$$
or equivalently,
\Eq{\label{S_omega} S^* \circ \hat{\omega}_L = - \hat{\omega}_L \circ S ,}
where $\hat{\omega}_L$ denotes the map $T(TQ) \to T^* (TQ)$ defined by
contraction with $\omega_L$.

Notice that $L$ being regular implies that the 2--form $\omega_L$ is
nondegenerate.  Consequently we can define its inverse bivector $\Lambda_L$ as,
\Eq{\label{lambda_l} \Lambda_L (\alpha , \beta ) = \omega_L (\hat{\omega}_L^{-1}
(\alpha ), \hat{\omega}_L^{-1} (\beta ) ), ~~~~~ \forall \alpha, \beta \in 
T^*(TQ)
.}

\medskip

Now for reasons that in principle have no relation with the
Lagrangian $L$, a submanifold $C\subset TQ$ will be selected, the
constraint submanifold of the problem.
If the submanifold $C$ is not of the form $TN$ with $N\subset Q$, or more
generally, if $C$ is not the total space of an integrable distribution $\sp D$
defined along a submanifold $N\subset Q$, we will say that the constraints are
non-holonomic.  In general
$\Gamma_L$ will not be tangent to $C$, this is, the Lagrangian system defined by
$L$ will evolve in time without keeping within the limits imposed by $C$.  If we
want the system to remain in $C$ then $\Gamma_L$ must be changed.   For that we
will assume that there is a set of ``forces'' $F\subset T^*(TQ)$ that allow
us to act upon the dynamical system $\Gamma_L$ and eventually to make it
to be confined to $C$.

The modified systems that we can obtain from $\Gamma_L$ by means of
the forces $F$ are given by the family of vector fields,
\Eq{\label{gamma}  \Gamma \in \Gamma_L + \Lambda_L (F)  ,}
where $\Lambda_L$ is the Poisson tensor defined by the Cartan 2--form
$\omega_L$, eq. (\ref{lambda_l}).   Notice
that Euler--Lagrange's equations (\ref{euler_lag}) in the presence of external
forces $f = f_i (q,v) dq^i$ are modified as
$$\lie{\Gamma} \Theta_L = dL + f ,$$
thus, spanning the Lie derivative, we will get that if $f\in F$, then $\Gamma$
verifies eq. (\ref{gamma}).
 
If we want the resulting system to be a system of ``mechanical'' type,
the system of ``forces'' will have to be given by horizontal
1--forms, i.e., $S^* (F) = 0$.  Then a simple computation shows that
$\Gamma$ is a {\sc sode} because $S(\Lambda_L (F)) = \Lambda_L (S^* (F)) = 0$
where we have used eqs. (\ref{S_omega}) and (\ref{lambda_l}).  

The system of forces $F$ will be obtained either by a detailed analysis of
the constraint submanifold or they will be given in an
independent way.   For instance, a large class of non-holonomic constraints are
originated by the interaction between the surfaces of different components of 
the
system.  In this category fall sliding, rolling and friction constraints,
which are supposed to create linear relations between the velocities of the
components of the system.  However, we can also imagine that our system
is subjected to the action of servomechanisms or other devices, that modify its
dynamical state in such a way that certain ``a priori'' given conditions 
are satisfied, for instance, limitations in the acceleration of the center of 
mass
of the system. The system of forces $F$ can be postulated from the non-holonomic
constraints, i.e., they will be supposed to be created by the constraints and
they will be supposed to have an explicit relation with them.  This is the case
of the so called Chetaev's conditions for non-holonomic constraints (see Section
5).  Some of these possibilities will be explored later on.   In this sense the
system of forces will be called reaction or control forces, depending if they
are derived from the constraints or are imposed externally.

Thus the Lagrangian non-holonomic constraint problem can be stated as
follows: given a regular Lagrangian $L$, a constraint submanifold
$C$ and a system of forces $F$, determine wether or not there exists a vector
field $\Gamma$ of the form given by eq. (\ref{gamma}) parallel to $C$.

\bigskip

This problem admits a simple generalization that is of interest in a
variety of situations.  We can replace the symplectic manifold $(TQ,
\omega_L)$ by a Poisson manifold $(P,\Lambda )$ and the dynamical
vector field $\Gamma_L$ by a Hamiltonian vector field $\Gamma_H =
\Lambda (dH)$.  We will consider a constraint submanifold $C \subset
P$ and, contrary to the situation discussed above, we can think that a
further restriction on the possible directions of the vector field
$\Gamma_H$ can be imposed, i.e., a field of allowed or permitted
directions $D$ along $C$ will be introduced\footnote{If we are discussing a
Lagrangian system, such restriction can be understood as a limitation on the
accelerations of the system.}.  The geometrical model for it will consists in a
vector subbundle $\rho\colon D\to C$ of $\tau_P
\colon TP \to P$ (see Fig. \ref{bundles}).  The reaction forces now will be
modelled by a subbundle $\eta\colon F \to C$ of $\pi_P\colon T^* P \to P$.  
Then,
the generalized non-holonomic problem can be stated as follows:  

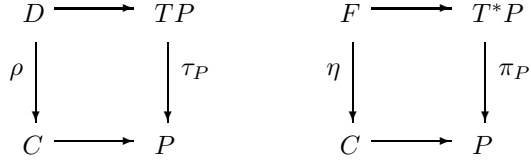
\begin{figure}

\begin{picture}(200, 50)(-90,0)
\put(0,0){$C$}
\put(50,0){$P$}
\put(0,50){$D$}
\put(50,50){$TP$}
\put(-5,30){$\rho$}
\put(60,30){$\tau_P$}
\put(12,5){\vector(1,0){30}}
\put(12,55){\vector(1,0){30}}
\put(5,42){\vector(0,-1){30}}
\put(55,42){\vector(0,-1){30}}

\put(120,0){$C$}
\put(170,0){$P$}
\put(120,50){$F$}
\put(170,50){$T^*P$}
\put(115,30){$\eta$}
\put(180,30){$\pi_P$}
\put(132,5){\vector(1,0){30}}
\put(132,55){\vector(1,0){30}}
\put(125,42){\vector(0,-1){30}}
\put(175,42){\vector(0,-1){30}}
\end{picture}

\caption{\label{bundles}  Non-holonomic constraints data, field of directions
and forces.}
\end{figure}

\bigskip

{\parindent 0cm {\bf Poisson non-holonomic constraint problem:}}  Given a
Hamiltonian vector field $\Gamma_H$ on a Poisson manifold $P$ with Poisson 
tensor
$\Lambda$, a constraint submanifold $C$, a field of permitted directions $D$ and 
a
system of control forces $F$, determine if there is a vector field $\Gamma$ of 
the
form \Eq{\label{gamma_h} \Gamma \in \Gamma_H + \Lambda (F) ,} 
contained in $D$.

\medskip
  
We should point it out that in both versions of the non-holonomic
constraint problem the sought vector field can exists or not, it can exists
only on a subset of the constraint submanifold, and even if it exists,
it is not necessarily unique.  It is clear that all kind of possibilities can
actually occur and it is not difficult to exhibit examples of them \cite{Me95}. 
The discussion in Sections 3 and 4 address the existence
and uniqueness of such dynamical vector fields. 
Particular situations will be also discussed where explicit conditions can be 
done
that will allow to give definite answers to them.

\section{Implicit differential equations}         

The main tool to analyze the general problems discussed above will be the
geometrical setting developed in \cite{Me95} to describe implicit differential
equations.

An implicit differential equation on the manifold $P$ is a submanifold $E\subset
TP$.  A solution of $E$ is any curve $\gamma \colon I \to P$, $I\subset \R$, 
such
that the tangent curve $(\gamma (t), \dot{\gamma}(t) ) \in E$ for all $t\in I$.  
The implicit differential equation will be said to be integrable at a point if
there exists a solution $\gamma$ of $E$ such that the tangent curve passes 
through
it.  The implicit differential equation will be said to be integrable if it is
integrable at all its points.   Integrability does not imply uniqueness.  The
integrable part of $E$ is the subset of all integrable points of $E$.  The
integrability problem consists in identifying such subset.

Denoting as before the canonical projection $TP \to P$ by $\tau_P$, a sufficient
condition for the integrability of $E$ is 
\Eq{\label{integr}  E \subset TC ,}
where $C = \tau_P (E)$, provided that the projection $\tau_P$ restricted to $E$ 
is
a submersion onto $C$.

\bigskip

{\parindent 0cm {\bf Extracting the integrable part of $E$:}}  A recursive
algorithm was presented in \cite{Me95} that allows to extract the integrable 
part
of an implicit differential equation $E$.  We shall define the submanifolds
\Eq{\label{level_0}  E_0 = E , ~~~~~ C_0 = C  ,}
and recursively for every $k \geq 1$,
\Eq{\label{level_k}  E_k = E_{k-1} \cap TC_{k-1} , ~~~~~ C_k = \tau_P (E_k) , }
then, eventually the recursive construction will stabilize in the sense that 
$E_k
= E_{k+1} = \cdots = E_\infty$, and $C_k = C_{k+1} = \cdots = C_\infty$.  It is
clear by construction that $E_\infty \subset TC_\infty$.   Then, provided that 
the
adequate regularity conditions are satisfied during the application of the
algorithm, the implicit differential equation $E_\infty$ will be integrable and 
it
will solve the integrability problem.

In what follows we will refer to the non-holonomic problem defined by the
Lagrangian $L$, the constraint submanifold $C$, the distribution $D$ and the 
system
of forces $F$, as the system $(L,C,D,F)$.

\section{Integrability of non-holonomic systems}
 
\subsection{The implicit differential equation associated to a system
with non-holonomic constraints}

The implicit differential equation associated to a non-holonomic system can be
described in the Lagrangian and/or the Hamiltonian formalism.

\medskip

{\parindent 0cm {\bf Lagrangian picture:}}   We will be given the data defining 
a
non-holonomic constrained Lagrangian system, i.e., a regular Lagrangian $L$, a
field of permitted directions $D \subset T_C (TQ)$ along the constraint 
submanifold
$C \subset TQ$ and a system of forces $F \subset T_C^* (TQ)$.  Then, the 
implicit
differential equation associated to $L, D, F$ will be defined by
\Eq{\label{lag_imp}  E = \alpha^{-1} (dL + F) \subset T(T^* Q) .}
Notice that $\tau_{T^*Q} (E) = C^*$ is the constraint submanifold considered in 
the
problem and it coincides with $T\sp F_L (C)$ where $\sp F_L$ denotes the
Legendre transformation defined by $L$ (see below).  It is also noticeable that
$dL + F$ defines an affine subbundle of $T^* (TQ)$ along
$C$.  The regularity of $L$ guarantees the transversality of the
submanifold $E$ with respect to the projection map $\tau_{T^*Q}$. 

\bigskip

{\parindent 0cm {\bf Hamiltonian picture:}}  As in the Lagrangian case, a
non-holonomic constrained Hamiltonian system is defined by a Hamiltonian 
function
$H$ on $T^*Q$, a permitted directions field $D\subset T_{C^*} (T^*Q)$ along the
submanifold $C^*\subset T^*Q$ and a system of forces $F\subset T_{C^*}^*
(T^*Q)$.  The implicit differential equation associated to the data $(H,C^*,
D,F)$ is defined by,
\Eq{\label{ham_imp} E = \beta^{-1} (dH + F) ,}
that projects onto $C^*$.

\subsection{Non-holonomic integrability algorithm}

If we apply the integrability algorithm given by eqs.
(\ref{level_0})-(\ref{level_k}) to the implicit differential
equation associated to a non-holonomic Lagrangian system given by 
eq. (\ref{lag_imp})\footnote{Notice that a similar algorithm can be applied
to non-holonomic Hamiltonian problems, eq. (\ref{ham_imp}).}, we will obtain,
$$\begin{array}{ll} E_0 = E = \alpha^{-1}(dL + F) ,  & C_0 = \tau_{T^*Q} (E) = 
C^*
, \\ E_1 = E_0 \cap TC_0 , & C_1 = \tau_{T^*Q} (E_1) , \ldots , \end{array}$$
and eventually $E_\infty \subset TC_\infty$.  However we must notice that the
stability of the algorithm does not imply automatically that $E_\infty$ is going
to define a vector field lying in $D$ and the integrable equation will not be in
general a solution of the non-holonomic problem.

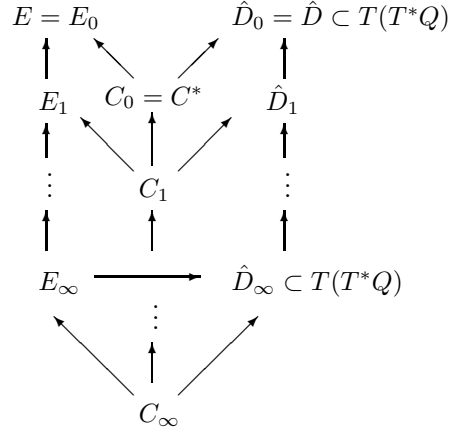
\begin{figure}

\begin{picture}(200, 140)(-120,0)
\put(25,0){$C_\infty$}
\put(-13,50){$E_\infty$}
\put(60,50){$\hat{D}_\infty \subset T(T^*Q)$}
\put(40,10){\vector(1,1){30}}
\put(23,10){\vector(-1,1){30}}
\put(8,55){\vector(1,0){40}}
\put(30,15){\vector(0,1){15}}
\put(30,35){$\vdots$}

\put(-10,130){\vector(0,1){15}}
\put(80,130){\vector(0,1){15}}
\put(-23,150){$E = E_0$}
\put(60,150){$\hat{D}_0 = \hat{D} \subset T(T^*Q)$}
\put(12,120){$C_0 = C^*$}
\put(30,97){\vector(0,1){20}}
\put(40,130){\vector(1,1){15}}
\put(23,130){\vector(-1,1){15}}

\put(-10,65){\vector(0,1){15}}
\put(80,65){\vector(0,1){15}}
\put(-10,85){$\vdots$}
\put(80,85){$\vdots$}
\put(-10,100){\vector(0,1){13}}
\put(80,100){\vector(0,1){13}}
\put(-13,118){$E_1$}
\put(73,118){$\hat{D}_1$}
\put(25,85){$C_1$}
\put(30,65){\vector(0,1){15}}
\put(40,95){\vector(1,1){20}}
\put(23,95){\vector(-1,1){20}}

\end{picture}

\caption{\label{inf} The non-holonomic integrability algorithm.}
\end{figure}

Then, the integrability algorithm needs a small adaptation to render the 
required outcome in the context of non-holonomic systems.
We need to translate the field of allowed directions to $T^*Q$.  The natural way
to do that is by means of the Legendre transformation $\sp F_L$ defined by $L$.
The derivative of this map $T\sp F_L \colon T(TQ) \to T (T^*Q)$ allows to
define the bundle of permitted directions on $T^*Q$ as
$$ \hat{D} = T\sp F_L (D) \subset T(T^*Q) ,$$
then we will construct an adapted integrability algorithm as follows,
$$ \begin{array}{lll}
E_0 = E, & C_0 = C^* , & \hat{D}_0 = \hat{D} , \\
E_1 = E_0 \cap \hat{D}_0, & C_1 = \tau_{T^*Q}(E_1) , & \hat{D}_1 =
\hat{D}_0\mid_{C_1} \cap TC_1 , \\
\cdots & \cdots & \cdots \\
E_k = E_{k-1} \cap \hat{D}_{k-1},  & C_k = \tau_{T^*Q}(E_{k}) , &
\hat{D}_k = \hat{D}_{k-1}\mid_{C_k} \cap TC_k , \\ 
\cdots & \cdots & \cdots ,
\end{array}$$
and then, eventually when the algorithm stops, we will obtain the
stable data $E_\infty$ , $\hat{D}_\infty$ , $C_\infty$ defining an
integrable implicit differential equation (see Fig. \ref{inf}).

Notice that $E_\infty \subset \hat{D}_\infty \subset TC_\infty$
which together with the appropriate regularity assumptions guarantees the
integrability of $E_\infty$ and its compatibility with the non--holonomic
constraints of the problem.  

\medskip

Obviously all kind of situations for $E_\infty$, $\hat{D}_\infty$, $C_\infty$
can happen.  The simplest nontrivial\footnote{Nontrivial here means that
$C$ is actually a submanifold of $TQ$ of codimension at least one and $F$ is a
subbundle of rank at least one.} situation occurs when $E_1 = E_\infty$,
$C^* = C_\infty$ and $\hat{D}_\infty = \hat{D}$.  Then, the algorithm stops in 
the
first step and,
\Eq{\label{compa} E_1 = E\cap \hat{D} = E \cap T\sp F_L (D) ,~~~~  C_1 =
\tau_{T^*Q} (E_1) = C^* ,~~~~  \hat{D}_1 = \hat{D}\cap TC^* = T\sp F_L (D)  .}
We will say that the regular Lagrangian $L$ and the non-holonomic constraints
$D$, $F$ are compatible if the conditions above, eq. (\ref{compa}), are met, 
i.e.,
the adapted integrability algorithm for them stops at the first step.  In this
case, the integrable implicit differential equation defined by them are the
submanifold of $T(T^*Q)$ given by,
\Eq{\label{eq_compa} E_\infty = \alpha^{-1} (dL + F) \cap T\sp F_L (D) ,}
and it gives the solution of the non-holonomic Lagrangian problem $(L,C,D,F)$.

\medskip

We must notice here that the outcome of the integrability algorithm does not
necessarily provide us with a {\sc sode}.  If we require that the solution must
be a {\sc sode}, then we must modify the algorithm again to incorporate this
fact.  This problem has been exhaustively discussed in the setting of
Lagrangians systems were it is known as the {\sc sode} problem for singular
Lagrangians.

\subsection{Compatibility conditions}

It is important to remark again that the previous characterization of
compatibility, eq. (\ref{compa}), does not imply the uniqueness of solutions
passing through its points.  This is a general feature of non-holonomic
constraints often encoutered in practical discussions.  It is also important to
notice again that even if the solutions were unique, i.e., the implicit
differential equation $E_\infty$ would define a {\it bona fide} differential
equation, they would not be necessarily solutions of a {\sc sode}.    The
following theorem gives sufficient conditions for the non-holonomic Lagrangian
problem $(L,C,D,F)$ to be compatible and, as a bonus, it is found that they also
guarantee the uniqueness and the second order character of it.

\begin{theorem}\label{comp}
Let $(L,C,D,F)$ be a non-holonomic Lagrangian problem.  If 
\Eq{\label{dim_cond} \rank (F) = \corank (D) ,}
and 
\Eq{\label{df}  D^0 \cap F^\perp = 0  ,} 
then the Lagrangian and the non-holonomic constraints are compatible.  Moreover,
there exists a unique {\sc sode} $\Gamma$ such that its graph is the integrable
implicit differential equation $\alpha^{-1}(dL + F)\cap T\sp F_L(D)$ associated
to it.
\end{theorem}

{\parindent 0cm {\bf Proof:}} 
The symbol $D^0$ denotes the annihilator of $D$, i.e.,
$$D^0 = \set{ a\in T_C^*(TQ) \mid \langle a, D\rangle = 0 } ,$$
and $F^\perp$ denotes the symplectic orthogonal to $F$ with respect to the
Poisson tensor $\Lambda_L$, that is,
$$F^\perp = \set{f \in T_C^* (TQ) \mid \Lambda_L (f,F) = 0 } .$$
The vector fields constructed from $\Gamma_L$ to solve the Lagrangian
non-holonomic problem have the form
$$\Gamma = \Gamma_L + \Lambda_L (f), ~~~~~~ f \in F .$$
If $\Gamma \in D$, then pairing the previous equation with elements in $D^0$, we
will get
$$\Lambda_L (a,f) = \langle \Gamma_L , a \rangle , ~~~~~\forall a\in D^0 .$$
Consider the map $\phi\colon F \to (D^0)^*$ defined by $\phi (f)(a) = \Lambda_L
(a,f)$, for all $a\in D^0$, whose kernel is $F \cap (D^0)^\perp$. 
Notice that $(F \cap (D^0)^\perp)^\perp = F^\perp + D^0$.  But
$\dim (F^\perp + D^0) = \dim F^\perp + \dim D^0$ because of eq. (\ref{df}), 
then,
$$\dim (F \cap (D^0)^\perp ) = 2n - (\dim F^\perp + \dim D^0 ) = 2n - (2n - 
\rank
F) - \corank D = 0 .$$
Then we conclude that $F \cap ( D^0)^\perp = 0$ and thus conditions
(\ref{dim_cond})-(\ref{df}) imply the injectivity of the map $\phi$.
Moreover, again because of $\rank (F) = \corank (D)$, the previous
map is surjective.  Hence the map $\phi$ is a bundle isomorphism and there 
exists
a unique element $f \in F$ such that $\phi (f) =
\langle \Gamma_L , . \rangle$.  Then there exists a unique vector field that 
gives
a solution of the non-holonomic Lagrangian problem, i.e., the submanifold $E$
cuts $T\sp F_L (D)$ along a submanifold which is the graph of the Legendre
transfom of the vector field $\Gamma$.  Moreover, because the forces $f$ a
horizontal 1--forms, the vector field $\Lambda_L (f)$ is vertical and $\Gamma$ 
is
a {\sc sode}.    \hfill $\Box$

\bigskip

{\parindent 0cm {\bf Remark:}}  It follows from the proof of the previous
theorem that the conditions for compatibility can
also be written as
$$ \rank (F^\perp ) = \corank (D^0), ~~~~ F \cap (D^0)^\perp = 0 ,$$
or they can be simultaneously encoded as
$$ T_C^* (TQ) = F^\perp \oplus D^0 .$$

\subsection{Local expressions}

Because of its practical interest we shall write the conditions in
Thm. \ref{comp} in local coordinates.

We will suppose that the submanifold $C$ is defined locally by the set of
functions 
$$\phi^a (q,v) = 0,  ~~~~~ a= 1, \ldots, r.$$
Then $\dim C = 2n -r$.  
The annihilator of the subbundle
$D\subset TC$ will verify $(TC)^0 \subset D^0$.  Thus, we will decompose $D^0 =
(TC)^0 \oplus K$.  We will suppose that $K$ is a subbundle with a local basis in
the chart $q^i, v^i$ given by the set of 1--forms 
$$\beta^b = \gamma_i^b dq^i + \beta_i^b dv^i,  ~~~~ b = 1, \ldots, s. $$  
A system of local 1--forms generating $(TC)^0$ is given by $d\phi^a$, $a= 1,
\ldots, r$.  Thus, $D^0$  will be locally generated by the set of 1-forms
$d\phi^a, \beta^b$, that will be collectively denoted by
$$\beta^C = \gamma_i^C dq^i + \beta_i^C dv^i,  ~~~~ C = 1, \ldots, r +
s . $$  
The corank of $D$ will then be $(r+s)$.  

We will now assume that the system of forces
$F$ will be defined by a subbundle of rank $r+s$, satisfying thus eq.
(\ref{dim_cond}), and it will have a system of local generating horizontal
1--forms $f^C = f_i^C dq^i$, $C = 1,\ldots, r+s$.  The local expression of the
Poisson tensor $\Lambda_L$ is given by
$$ \Lambda_L = W^{ij} \dd{q^i}\wedge \dd{v^i} + M^{ij} \dd{v^i} \wedge \dd{v^j} 
.$$
The intersection condition (\ref{df}) is equivalent to the statement that
$\Lambda_L$ defines a nondegenerate pairing between $F$ and $D^0$, i.e., that 
the
matrix 
$$ M^{BC} = W^{ij}(q,v) f_i^B (q,v) \beta_j^C (q,v) ,$$ 
is nondegenerate.  In the particular case that $D = TC$, i.e., there are no
limitation to the permitted directions for the vector field, then the set of
1-forms spanning $D^0$ are given by $d\phi^a$, $a= 1,\ldots, r$.  Then, the
compatibility condition becomes,
$$ 0 \neq \det M^{ab} = W^{ij}(q,v) f_i^a (q,v) \pd{\phi^b}{v^j} .$$

\bigskip

{\parindent 0cm {\bf Linear non-holonomic constraints:}}

The situation that usually is dealt with is when the submanifold $C$ is an
affine subbundle $A$ of $TQ$, i.e., locally, $C$ is defined by local functions,
$$ \phi^a (q,v) = \mu^a_i (q) v^i + \mu^a_0 (q) = 0 ,~~~~~~  a= 1, \ldots, r.$$ 
The tangent submanifold $TA \subset T_A(TQ)$ is defined by the set of 1-forms
$d\phi^a$, i.e., $(TA)^0 = \span \set{d\phi^a \mid a = 1,\ldots, r}$.  Hence,
because 
$$ d\phi^a = \mu_i^a (q) dv^i + \left( v^i \pd{\mu_i^a}{q^j} + \pd{\mu_0^a}{q^j}
\right) dq^j ,$$
namely, 
$$ \pd{\phi^a}{v^i} = \mu_i^a (q) ,$$
and the compatibility condition becomes,
\Eq{\label{mab} \det M^{ab} = \det W^{ij} (q,v) f_i^a (q,v) \mu_j^b (q) \neq 0 
.}
If, in adition we accept Chetaev's forces (see next Section), i.e, we define,
\Eq{\label{lin_for} F = S^* (TA)^0 ,}
the subbundle $F$ will be generated by the 1-forms $\mu^a = \mu_i^a (q)
dq^i$ because $S^* (d\phi^a) = \mu_i^a (q) dq^i$, and the compatibility
condition will be simply given by
\Eq{\label{lin_comp} \det  M^{ab} = \det W^{ij} (q,v) \mu_i^a (q) \mu_j^b (q)
\neq 0 ,} 
reproducing results in \cite{Le96}.   Notice that if $L$ is a
mechanical Lagrangian $L = T - V$, where $T$ is the kinetic energy corresponding
to a Riemannian metric $g$, then, 
$$ M^{ab} = g^{ij} \mu_i^a (q) \mu_j^b (q) ,$$
which is obviously invertible.

\section{Comparison with previous results}

We have already seen in the previous section, that particular choices of the set
of reaction forces $F$, for instance those given in eq. (\ref{lin_for}), give
special expressions for the compatibility conditions discussed in Section 4
before, Thm. \ref{comp}.   A special choice for the reaction forces is given by
Chetaev's conditions. In intrinsic terms the system of forces corresponding to a
non-holonomic constraint submanifold
$C\subset TQ$ under Chetaev's conditions is given by the subbundle
(see also \cite{Ma91}),
\Eq{\label{chetaev}  F = S^* ((TC)^0) .}
More generally, if we consider a bundle of permitted directions $D\to C$, then
Chetaev's bundle of reaction forces will be given by (see also \cite{Le96}),
\Eq{\label{gen_chetaev} F = S^* (D^0) .}
Following \cite{Fa72} we will say that $D$ is admissible if $S^*$ is injective
when restricted to $D^0$, i.e., $D^0$ does not contains horizontal 1--forms. 
This implies that the rank of the Chetaev bundle coincides with the rank of
$D^0$, i.e., $\rank F = \corank D$.  Hence, Faddeev's admissibility condition is
a particular instance of the condition (\ref{dim_cond}) in Thm. \ref{comp}.  

The Hessian of the Lagrangian $L$ can be defined as the symmetric
$(0,2)$-tensor\footnote{Notice that $H_L$ is symmetric because of eq.
(\ref{S_omega})},
$$H_L (\alpha , \beta ) = \Lambda_L (S^*(\alpha ), \beta ) , ~~~~~ \alpha,
\beta \in T^*(TQ).$$ 
Then, we will say that $L$ is definite if the Hessian $H_L$
is definite as a symmetric tensor.   Then, if $L$ is definite, $H_L$
is nondegenerate when restricted to the subspace $D^0$, but $H_L
(D^0, D^0) =
\Lambda_L (S^*(D^0), D^0) = \Lambda_L (F, D^0)$, and $\Lambda_L$ is 
nondegenerate
on the pair $F$, $D^0$, i.e., $F^\perp \cap D^0 = 0$.  Then if $L$ is definite
(``normal'' according to the terminology in \cite{Ma90}) this implies condition
(\ref{df}).  Then Thm. \ref{comp} implies the main result in \cite{Fa72},
\cite{Le96}.

\begin{theorem}\label{Faddeev}
Let $L$ be a definite Lagrangian and a non-holonomic constraint defined by the
permitted field of directions $D\to C$.  If $D$ is admissible and the reaction
forces are given by the Chetaev's bundle $F= S^* (D^0)$, then there exists a
unique {\sc sode} $\Gamma$ solution of the non-holonomic Lagragian problem
defined by $L$ and $D\to C$.
\end{theorem}

\bigskip

A different condition is used in \cite{Ba93} and \cite{Le96} to find a solution
of the Lagrangian non-holonomic problem.  Now we will assume that a submanifold
$C\subset TQ$ is given and $D = TC$.  Let $\sp S_C$ be the distribution 
\Eq{\label{leon}  \sp S_C = S (TC^\perp )  .}
A Lagrangian system $L$ with a non-holonomic constraint is said to be regular if
\Eq{\label{reg_int} \sp S_C \cap TC = 0 .}  
The following theorem is again a
particular instance of Thm. \ref{comp}.

\begin{theorem}
If $(L, C)$ is a regular non-holonomic Lagrangian system, then it has a unique
solution.  In addition this solution is a {\sc sode}.
\end{theorem}

{\parindent 0cm {\bf Proof:}}  The analysis of the regularity condition $\sp S_C
\cap TC = 0$ leads to,
\Eq{\label{condition} 0 = \hat{\omega}_L (\sp S_C \cap TC ) = \hat{\omega}_L
(\sp S_C) \cap \hat{\omega}_L (TC) = \hat{\omega}_L (S(TC^\perp )) \cap
\hat{\omega}_L (((TC)^0)^0) ,}
where as before $\hat{\omega}_L$ is the natural bundle map $T(TQ) \to T^*(TQ)$
defined by contraction with the 2--form $\omega_L$.  Then it is obvious that
$\hat{\omega}_L (TC^\perp ) = TC^0$.  In fact, $\alpha \in \hat{\omega}_L
(TC^\perp )$ iff there is $u\in TC^\perp$ such that $\hat{\omega}_L (u) =
\alpha$,  but $\omega_L (u,TC) = 0$, then $\alpha \in TC^0$ and conversely.  
Then,
continuing with the computation in eq. (\ref{condition})
we get,
$$\hat{\omega}_L (S(TC^\perp )) \cap\,
\hat{\omega}_L (((TC)^0)^0) = S^* (\hat{\omega}_L (TC^\perp )) \cap (TC^0)^\perp 
=
S^* (TC^0) \cap (TC^0)^\perp = F\cap (D^0)^\perp ,$$
with $F= S^*(TC^0)$ and $ D= TC$,
and again we are in the conditions of Thm. \ref{comp}. \hfill$\Box$

\bigskip

As it was pointed before, a similar approach to this paper was adopted by S.
Benenti in \cite{Be87}.  There, the problem of linear non-holonomic Lagrangian 
or
Hamiltonian systems was addressed and the compatibility conditions exhibited
there (Props. 2 and 3) are equivalent to the regularity of $L$ and eq.
(\ref{lin_comp}).  The nonlinear case is also discussed and the conditions in
Prop. 4 of \cite{Be87} are equivalent respectively to the regularity of the
Lagrangian, the rank condition in Thm. \ref{comp}, and the regularity of the
matrix $M^{ab}$ in eq. (\ref{mab}) plus Chetaev's conditions;  hence the
conclusions there follow from Thm. \ref{comp}. 

\medskip

To end this section we will comment on the characterization given by C.M. Marle
in \cite{Ma90}.  The subbundles $W$ (called the projection bundle) and $TD$, the
tangent bundle of the Hamiltonian constraint submanifold, are $W = T\sp F_L
(\sp S_C)$ and $D = C^*$, using the notation in this paper.  Then, $W\oplus TD$
is the translation to $T^*Q$ of the regularity condition eq. (\ref{reg_int}). 
Then, the existence and uniqueness of solutions stated in
\cite{Ma90} (Prop. 2.15 and Thm. 2.16) are again particular instances of the
previous results.

\section{Examples and applications}

\subsection{Disk rolling on a surface}

As a simple example we will discuss first the a disk rolling vertically on a
rough surface.   The disk of mass $M$ and radius $R$ will be described by the
coordinates $x,y$ of the contact point with the surface, the angle
$\theta$ defined by the plane containing the disk and a fixed plane normal to
the surface, and the angle $\phi$ parametrizing the position of the disk with
respect to its center (see also Fig. \ref{appell_0}).   The configuration space 
of
the system will be $Q = \R^2 \times \T^2$.  The Lagrangian describing the free
system is,
\Eq{\label{disk_rol} L = \mitad M(\dot{x}^2 + \dot{y}^2) + \mitad I_1
\dot{\theta}^2 + \mitad I_2 \dot{\phi}^2 ,}
whith $I_1, I_2$ the corresponding moments of inertia.
The constraint submanifold $C\subset TQ$ is given by the rolling conditions,
\Eq{\label{disk_cons} \Psi_1 = \dot{x} - R \cos \theta \dot{\phi} = 0; ~~~~~~ 
\Psi_2 = \dot{y} - R \sin \theta \dot{\phi} = 0 .}
In the cotangent bundle $T^*Q = T^*\R^2 \times T^* \T^2$ we introduce 
coordinates
$(x,y,\theta, \phi , p_x, p_y, p_\theta , p_\phi )$, and finally in $T(T^*Q)$ we
will have coordinates $(x,y,\theta, \phi , p_x, p_y, p_\theta ,
p_\phi; \dot{x}, \dot{y}, \dot{\theta}, \dot{\phi} , \dot{p}_x, \dot{p}_y,
\dot{p}_\theta , \dot{p}_\phi )$.  
The annihilator of $TC$ on $T_C^*(TQ)$ will be spanned by $d\Psi_1$, $d\Psi_2$.
Using Chetaev's forces, we will have, $F = S^* (TC)^0$,
$$ F = \set{ \lambda_1 (dx - R\cos \theta \,d\phi) + \lambda_2 (dy - R\sin 
\theta
\, d\phi) \mid \lambda_1, \lambda_2 \in \R } .$$
Thus, the affine subbundle $dL + F \subset T_C^*(TQ)$ is given by,
\begin{eqnarray*}
dL + F &=& \set{ M \dot{x}d\dot{x} + M \dot{y}d\dot{y} + I_1 \dot{\phi}
d\dot{\phi} + I_2 \dot{\theta} d\dot{\theta} + \\
&& + \lambda_1 (dx - R\cos \theta\,
d\phi) + \lambda_2 (dy - R\sin \theta\, d\phi) \mid \lambda_1\lambda_2 \in \R }.
\end{eqnarray*} 
The map $\alpha \colon T(T^*Q) \to T^*(TQ)$ is defined in coordinates $(q^i,
p_i; \dot{q}^i, \dot{p_i})$ for $T(T^*Q)$ and $(q^i,v^i; r_i, s_i)$ by
$$s_i\circ \alpha = p_i, ~~~ r_i\circ \alpha = \dot{p}_i , ~~~~~ v^i \circ
\alpha = \dot{q}^i , ~~~~~ q^i \circ \alpha = q^i ,$$
or equivalently, $\alpha (q^i,p_i,\dot{q}^i, \dot{p}_i) = (q^i,\dot{q}^i,
\dot{p}_i, p_i)$.  Then, the submanifold $E= \alpha^{-1}(dL + F)$ is given by 
the
equations,
\begin{eqnarray*}
 E &=& \set{ (x,y,\theta, \phi, p_x,p_y,p_\theta, p_\phi;\dot{x}, \dot{y},
\dot{\theta}, \dot{\phi}, \dot{p}_x, \dot{p}_y, \dot{p}_\theta, \dot{p}_\phi)
\in T(T^*Q) \mid \\ 
&& \dot{p}_x = \lambda_1, \dot{p}_y = \lambda_2, \dot{p}_\phi = -\lambda_1 R 
\cos
\theta -\lambda_2 R \sin \theta,
\dot{p}_\theta = 0, p_x = M\dot{x}, p_y = M\dot{y}, \\
&& p_\phi = I_1\dot{\phi}, p_\theta = I_2\dot{\theta}, \Psi_1 = 0, \Psi_2 = 0 }.
\end{eqnarray*}
 The Legendre transformation $\sp F_L \colon TQ \to T^*Q$ is given by 
$$ p_x = M\dot{x}, p_y = M\dot{y}, p_\phi = I_1\dot{\phi}, p_\theta =
I_2\dot{\theta} ,$$ 
and maps the submanifold $C$ into the submanifold $C^*\subset T^*Q$ given by
$$ C^* = \set{(x,y,\theta,\phi ;p_x,p_y, p_\theta, p_\phi )\in T^*Q \mid I_1 p_x 
=
mR\cos \theta\, p_\phi, I_1 p_y = mR\sin \theta\, p_\phi} .$$
It is clear that $\tau_{T^*Q} (E) = C^*$, and thus $E\cap TC^* = E_1$ and the
integrability algorithm stops\footnote{This was known in advance because $L$
is definite and the constraints are admissible.}.  Computing the intersection of
$TC^*$ with $E$ we obtain,
$$ \lambda_1 = -\frac{MR\sin \theta}{I_1I_2} p_\phi p_\theta, ~~~~~ 
\lambda_2 = -\frac{MR\cos \theta}{I_1I_2} p_\phi p_\theta, ~~~~~ \dot{p}_\theta
= 0, ~~~ \dot{p}_\phi = 0 .$$
In this form we will obtain the following vector field $\Gamma$ defined on 
$C^*$,
$$ \Gamma = \frac{R\cos \theta}{I_1}p_\phi \frac{\partial}{\partial x} + 
\frac{R\sin \theta}{I_1} p_\phi \frac{\partial}{\partial y} + 
\frac{p_\phi}{I_1} \frac{\partial}{\partial \phi} + 
\frac{p_\theta}{I_2} \frac{\partial}{\partial \theta} -  
\frac{MR\sin \theta}{I_1I_2} p_\phi p_\theta \frac{\partial}{\partial p_x} +
\frac{MR\cos \theta}{I_1I_2} p_\phi p_\theta \frac{\partial}{\partial p_y} ,$$
which is mapped by $\sp F_L^{-1}$ into a {\sc sode},  or equivalently the
following set of equations of motion,
$$\begin{array}{llll}
\dot{x} = \frac{R\cos \theta}{I_1} p_\phi, & \dot{y} = \frac{R\sin \theta}{I_1}
p_\phi ,& \dot{\phi} = \frac{p_\phi}{I_1}, & \dot{\theta} = \frac{p_\theta}{I_2}
,\\  & & & \\ 
\dot{p}_x = - \frac{MR\sin \theta}{I_1I_2} p_\phi p_\theta , &
\dot{p}_y = \frac{MR\cos \theta}{I_1I_2} p_\phi p_\theta , &
\dot{p}_\phi = 0, & \dot{p}_\theta = 0 , \end{array} $$
whose solutions are found immediately.

The results thus obtained are in full agreement with the solutions obtained in 
any
elementary course in Analytical Mechanics and illustrate the predictions of Thm.
\ref{Faddeev}.  

\subsection{Appell's machine}

We will discuss now the well--known Appell's machine (see Figure
\ref{appell_0}). 

\begin{figure}
\vspace{5cm}
\caption{\label{appell_0}  Appell's machine with $\rho = 0$.}
\end{figure}

{\parindent 0cm $\rho = 0$}: It consists in a disk of radius $R$ with a small
cylinder rigidily attached to it of radius $r$ with total mass $M$.  A rope
passes through the small cylinder and hangs a mass $m$ on the other extreme of
a vertical frame.  The frame is of negligible mass and the disk rolls over the
surface.  The coordinates of the contact point of the main wheel will be
$(x, y)$ and the coordinates of the small mass $m$ will be thereby $(x,y,z)$.  
The
angle defined by the horizontal axis of the wheel with the $0x$ axis will be
denoted by $\theta$ and the angular position of the wheel will be denoted by
$\phi$ as in the example before.  The configuration space being $Q = \R^3 \times
\T^2$ and the tangent bundle becomes the 10-dimensional manifold $TQ = T\R^3
\times T\T^2$.  The Lagrangian of the system will be given by
$$ L = \mitad  M (\dot{x}^2 + \dot{y}^2 ) + \mitad m (\dot{x}^2 + \dot{y}^2 +
\dot{z}^2 ) + \mitad I_1 \dot{\phi}^2 + \mitad I_2 \dot{\theta}^2 + mgz .$$ 

The rolling conditions between the disk
and the floor and the nonsliding condition between the rope and the cylinder
define the constraint submanifold $C\subset TQ$.   The 7-dimensional submanifold
$C$ is then  characterized by the equations,
$$ \Psi_1 = \dot{x} - R\cos \theta \dot{\phi}=0, ~~~ \Psi_2 = \dot{y} - R\sin
\theta \dot{\phi} = 0, ~~~ \Psi_3 = r\dot{\phi} - \dot{z}=0 .$$ 
The non-holonomic constraints are linear and the system can be considered to be
of Chetaev's type with $F = S^*(TC)^0$. 
The Legendre transformation of $C$ gives the submanifold $C^*$,
\begin{eqnarray*} C^* &=& \set{ (x,y,z,\theta, \phi, p_x, p_y,p_z, p_\theta,
p_\phi) \in T^*Q \mid  I_1 p_x =
(M+m)R \cos \theta\, p_\phi, \\
&& I_1 p_y = (M+m)R \sin \theta \, p_\phi, I_1 p_z = mR p_\phi } .
\end{eqnarray*}

 Repeating the computations of the
previous section we will arrive to the following equations of motion on the
submanifold $C^*\subset T^*Q$,

\begin{eqnarray}
&& \dot{p}_x = \lambda_1, ~~~~~~ \dot{p}_y = \lambda_2, ~~~~~~ \dot{p}_z =
\lambda_3 - mg, \\
&& \dot{p}_\phi = - \lambda_1 R\cos \theta - \lambda_2 R \sin \theta -
\lambda_3 r , ~~~~~ \dot{p}_\theta = 0 .
\end{eqnarray}
Thus, solving the Lagrange's multipliers, $\lambda_1, \lambda_2, \lambda_3$,
i.e., computing $E\cap TC^*$, we get, 
$$ \ddot{\phi} = -\frac{mgr}{I_1 + (M+m)R^2 + mr^2}, ~~~~~~~~ \ddot{\theta} = 0
,$$
which is integrated immediately.
Two comments are pertinent here. Notice that the last equation is simply the
conservation of the $z$ component of the angular momentum, which is obvious from
the analysis of the forces acting upon the system.  Secondly, we know in
advance that the system is compatible and has a unique and well defined solution
without having to consider it as a limit case $\rho \to 0$ of a ``well posed''
system.  In fact we will see immediately that the case
$\rho \neq 0$  (see Fig. \ref{appell}) has  extra difficulties because of the
nonconservation of angular momentum.  

\bigskip

{\parindent 0cm $\rho \neq 0$}:
The system now consists as before on a disk an a small cylinder rigidily 
attached
to it.  A rope passes through the small cylinder and hangs a mass $m$ on the 
other
extreme of a frame of length $\rho$.  The frame is of negligible mass and slides
without friction.  Now the coordinates of the contact point of the main wheel 
will
be $(x_D, y_D)$ and the coordinates of the mass $m$ will be $(x,y,z)$.  The 
angle
defined by the horizontal axis of the wheel with the $0x$ axis will be denoted
again by
$\theta$ and the angular position of the wheel will be denoted by $\phi$.
The Lagrangian of the system will be given now by
$$ L = \mitad  M (\dot{x}_D^2 + \dot{y}_D^2 ) + \mitad m (\dot{x}^2 + \dot{y}^2 
+
\dot{z}^2 ) + \mitad I_1 \dot{\phi}^2 + \mitad I_2 \dot{\theta}^2 - mg z .$$ 
The geometry of the system imposes that $x - x_D = \rho \cos \theta$ and $y- y_D 
=
\rho \sin \theta$, (that can be understood as holonomic constraints).  We can
eliminate them from the Lagrangian $L$ and we get a system on $Q = \R^3 \times
\T^2$ as in the situation before.  The constraints now are given by,
$$ \Psi_1 = \dot{x}_D - R\cos \theta \dot{\phi}=0, ~~~ \Psi_2 = \dot{y}_D - 
R\sin
\theta \dot{\phi} = 0, ~~~ \Psi_3 = r\dot{\phi} - \dot{z}=0 .$$ 

\begin{figure}
\vspace{5cm}
\caption{\label{appell}  Appell's machine with $\rho \neq 0$.}
\end{figure}

Repeating the computations and using the Legendre transform to write now the
equations of motion directly on $C \subset TQ$ we will obtain,
\begin{eqnarray*}
&& (M+m) \ddot{x} + M\rho \dot{\theta}^2 \cos \theta + M\rho \ddot{\theta}
\sin\theta = \lambda_1, 
\\ && (M+m) \ddot{y} + M\rho \dot{\theta}^2 \sin \theta 
- M\rho \ddot{\theta} \cos \theta  = \lambda_2, 
\\ && m\ddot{z} + mg = - \lambda_3 \\
&& I_1 \ddot{\phi} = -\lambda_1 R \cos \theta - \lambda_2 R\sin \theta +
\lambda_3 r , 
\\ && (I_2 - M\rho^2)\ddot{\theta} - M\rho R \dot{\theta}\dot{\phi} =
\lambda_1\rho \sin \theta - \lambda_2 \rho \cos \theta
\end{eqnarray*}
Eliminating the Lagrange's multipliers, we obtain $\phi$ and $\theta$
variables,
\begin{eqnarray}
&& (I_2 + (M+m)\rho^2)\ddot{\theta} + \rho R m \dot{\theta} \dot{\phi} =
0, \label{dot_theta} 
\\ && (I_1 + (m + M)R^2 + m r^2) \ddot{\phi} - MR\rho \dot{\theta}^2 =
-mrg . \label{dot_phi}
\end{eqnarray}
Eq. (\ref{dot_theta}) shows that there is no conservation of the
angular momentum, the reason being that for the mass $m$ to move rigidily with
the frame, this has to act on it with a horizontal force.  The previous 
equations,
(\ref{dot_phi}), (\ref{dot_theta}), can be solved and we
obtain $\phi (t)$ implicitely from the integral,
$$\frac{1}{\sqrt{2}} \int_{\phi_0}^\phi \frac{d\phi}{\sqrt{ E - mgr\phi + 
\frac{I_2 + (M+m) \rho^2}{I_1 + (M+m)R^2 + mr^2} \exp (-\frac{2MR\rho}{I_2 -
(M+m)\rho^2}(\phi -\phi_0)}}= t - t_0,$$ and
$$ \theta (t) = \theta_0 + \int_{t_0}^t e^{-\frac{2MR\rho}{I_2 - 
(M+m)\rho^2}(\phi
(t) -\phi_0)} dt .$$

\subsection{A variation of Benenti's problem}

Finally we will consider some variants of an example proposed by Benenti
\cite{Be87} (see Fig. \ref{benenti}).

First we will solve the problem of two point massess forced to move on a plane 
with
parallel velocities.  The configuration space will be $Q = \R^2 \times \R^2$ and
we will denote by $(x_1,y_1)$ the position of the particle of mass $m_1$ and by
$(x_2,y_2)$ the position for the particle of mass $m_2$.  The
constraint on the velocities is given by the function on the tangent bundle
$TQ$,
\Eq{\label{paral} \Psi = \dot{x}_1 \dot{y}_2  - \dot{x}_2 \dot{y}_1 .}
The Lagrangian of the system is,
$$ L = \mitad m_1 (\dot{x}_1^2 + \dot{y}_1^2 ) + \mitad m_2 (\dot{x}_2^2 +
\dot{y}_2^2) .$$ 
The non-holonomic constraint $\Psi$ is a genuine non-linear
non-holomic constraint.   If we solve the system with Chetaev's forces
(\ref{chetaev}) we will obtain the following system of equations on $TQ$,
$$ m_1 \ddot{x}_1 = \lambda \dot{y}_2, ~~~ m_1 \ddot{y}_1 = -\lambda
\dot{x}_2, ~~~ m_2 \ddot{x}_2 = -\lambda \dot{y}_1, ~~~~ m_2 \ddot{y}_2 = 
\lambda
\dot{x}_1 ,$$
thus derivating eq. (\ref{paral}), and substituting on it, we get,
$$ \left( \frac{\dot{x}_2^2 + \dot{y}_2^2 }{m_1} + \frac{\dot{x}_1^2 +
\dot{y}_1^2}{m_2} \right) \lambda = 0 .$$   
Hence, $\lambda = 0$ and  Chetaev's forces vanish.  The solution of the system
is simply free motion of the two particles with initial parallel velocities.  
Notice that Chetaev's forces have the form $f_1 = \lambda (\dot{y}_2 dx_1 -
\dot{x}_2 dy_1)$, $f_2 = \lambda (-\dot{y}_1 dx_2 +\dot{x}_1 dy_2) $ where $f_i$
is the force acting on the particle $i$, thus the only solution leaving the
submanifold $\Psi = 0$ invariant is obtained with $\lambda = 0$.  However if we
choose the system of forces 
$$F =  \set{ \lambda (m_1f_1 dx_1 + m_1f_2 dy_1 + m_2f_1
dx_2 + m_2f_2 dy_2) \mid \lambda\in \R } ,$$ 
then we will have the equations,
$$ \ddot{x}_1 = \lambda f_1, ~~~ \ddot{y}_1 = \lambda f_2, ~~~
\ddot{x}_2 = \lambda f_1, ~~~~ \ddot{y}_2 = \lambda f_2 ,$$
thus if the initial velocities were parallel, the corresponding motions will be 
a solution for arbitrary $\lambda$.  Thus, in this particular situation there 
are
non unique solution of the problem.

\begin{figure}
\vspace{5cm}
\caption{\label{benenti} Benenti's system. }
\end{figure}

The problem can be modified substituting the point masses by disks rolling
without sliding on the plane.  Then we will be considering two copies of the
system discussed in Section 6.1., eqs. (\ref{disk_rol}), (\ref{disk_cons}), with 
the
additional constraint,
\Eq{\label{psi_5} \Psi_5 = {\mbox{\bf v}}_1\wedge  {\mbox{\bf v}}_2 ,}
where ${\mbox{\bf v}}_i$ denotes the velocity of the center of mass of the $i$th
disk.  Replacing the constraints given by the rolling conditions, eq.
(\ref{disk_cons}), on eq. (\ref{psi_5}), we will obtain,
$$ R_1 R_2 \sin (\theta_2 - \theta_1 ) \dot{\phi}_1 \dot{\phi}_2 = 0 .$$
Then, either one of the disks is still ($\dot{\phi}_i = 0$) and the angles
$\theta_i$ are arbitrary, or if the disks are rolling ($\dot{\phi}_i \neq 0$), 
then
$\theta_1 = \theta_2 = \theta$ and the nonlinear constraint $\Psi_5$ is 
redundant. 
Then the two disks roll independently and freely keeping parallel directions. 

If we remove the rolling conditions and we keep only the parallelism condition
$\Psi_5 =0$, then we will obtain again the solutions of the point masses problem
before, i.e., free motion with initial parallel velocities.  Under this 
condition,
the disks slide without rolling.  It is obvious that using non-Chetaev forces, 
other
solutions can be found.

\section{Conclusions and outlook}

We have reviewed the theory of Lagrangian systems with non-holonomic constraints
from the viewpoint of implicit differential equations in the realm of
symplectic and tangent bundle geometry.   From the beginning we have
deliberately separated the non-holonomic constraints from the forces or controls
that we can use to force the system to satisfy them, emphasizing in this way
that non-Chetaev's systems can be included in this picture.  In fact, we have
found general conditions that guarantee the existence and uniqueness of 
solutions
for this generalized non-holonomic problem.  We have especialized this general
compatibility conditions to different particular situations already discussed in
the literature.    

The discussion has been
made in the autonomous case but it is obvious that it can be extended to the
time--dependent setting easily (see for instance \cite{Ra94}, \cite{Let},
\cite{Gi92}).  Some of the beautiful aspects of Tulczyjew's triple are lost in
a time--dependent setting and this justifies to keep the discussion at the
autonomous level.   The usual extended phase space trick can be used to include
time as an additional variable and proceed as it has been done here.  An
alternative path would be to extend Tulcyjew's triple to the setting of
cosymplectic/contact geometry to recover the same geometrical setting without
introducing spurious variables.

Subtler is the problem of extending the theory of non-holonomic constraints to
singular Lagrangians.  This is a significant extension of the theory because
physical Lagrangians are very often singular.   Singular lagrangians introduce
their own constraints, that arise as integrability conditions of an implicit
differential equation.   A simultaneous analysis of Lagrangian constraints and
non-holonomic constraints is needed to obtain the equations of motion of the
theory.  This problem has been addressed in \cite{LeD} and \cite{Di96}.

An important aspect of the theory of non-holonomic constraints is its
relation with Quantum Mechanics.   Apart from a paper by R.J. Eden \cite{Ed51},
there has been little attention to the quantization of systems with 
non-holonomic
constraints (see also \cite{Pi89}).   One reason for this is that fundamental
theories do not include non-holonomic constraints because 
non-holonomic constraints are phenomenological models that offer a reasonable
description for the (often unknown) true fundamental theory that describes the
interaction between the surfaces of the system in contact, or other interactions
present in the system which cannot be described at the same fundamental level.  
Some
aspects concerning the Hamiltonian structure of non-holonomic systems and its
quantization are being discussed in \cite{Di96}.

\bigskip

{\parindent 0cm{\bf Ackwnoledgements}}  The authors AI, MdL and DMD, wish to
ackwnoledge the partial financial support provided by CICYT under the
programmes PB92-0197 and PB94-0106 respectively, as well as the NATO 
collaborative
research grant 940195.

{\parindent 0cm

{\sc A. Ibort} 

{\it Depto. de F{\'\i}sica Te\'orica, Univ. Complutense de Madrid,
28040 Madrid, Spain.} 

\bigskip

{\sc M. de Le\'on} 

{\it Inst. de Matem\'aticas y F{\'\i}sica Fundamental, CSIC, Serrano 123, 28006
Madrid, Spain.}

\bigskip

{\sc G. Marmo} 

{\it Dipto. di Scienze Fisiche, Univ. di Napoli, Mostra
d'Oltremare, Pad. 19, 80125 Napoli, Italy.}

\bigskip

{\sc D. Mart{\'\i}n de Diego} 

{\it Dpto. de Econom{\'\i}a Aplicada Cuantitativa, UNED,
28040 Madrid, Spain.} }

\end{document}